\documentclass[12pt]{article}

\usepackage{amssymb,amsmath,amsfonts,authblk,booktabs,blindtext, eurosym,geometry,ulem,graphicx,caption,color,setspace,sectsty,comment,footmisc,caption,pdflscape,subfigure,array,hyperref,float, multirow, pbox, rotating, adjustbox, caption, regexpatch, siunitx, setspace,tabularx}
\usepackage[english]{babel}
\usepackage[flushleft]{threeparttable}
\usepackage{array}
\usepackage{ragged2e}
\usepackage{multicol}
\usepackage{verbatim} 
\usepackage{graphicx}
\usepackage{adjustbox}
\usepackage{tabularx}
\usepackage{xurl}
\usepackage[colorinlistoftodos]{todonotes}
\usepackage{atbegshi}
\AtBeginDocument{\AtBeginShipoutNext{\AtBeginShipoutDiscard}}

\definecolor{darkblue}{rgb}{0.0,0.0,0.5}

\newcolumntype{P}[1]{>{\RaggedRight\hspace{0pt}}p{#1}}

\hypersetup{
	colorlinks= true,
	linkcolor = darkblue,
	urlcolor  = darkblue,
	citecolor = darkblue,
	anchorcolor = blue
}
\usepackage{csquotes}
\usepackage[style = apa, sorting = nyt,
			natbib = true,
			doi = false,
			isbn = false]{biblatex}
\begin{document} 

\title{An Event Study of the Ethereum Transition to Proof-of-Stake}
\author[1]{Elie Kapengut}
\author[2]{Bruce Mizrach}\thanks{Correspondence: Department of Economics, Rutgers University, 75 Hamilton Street, New Brunswick, NJ 08901 USA. email:
mizrach@econ.rutgers.edu, (908) 913-0253 (voice) and (425) 795-9942 (fax).}
\affil[1]{\small Rutgers University, New Brunswick, NJ USA}

\affil[2]{\small Department of Economics, Rutgers University, New Brunswick, NJ USA}
\date{First Draft: October 13, 2022 \\ Revised: February 25, 2023}

\begin{titlepage}
\maketitle
\begin{abstract}
\noindent On September 15, 2022, the Ethereum network adopted a proof-of-stake (PoS) consensus mechanism. We study the impact on the network and competing platforms in a two month event window around the Beacon chain merge. We find that the transition to PoS has reduced energy consumption by 99.98\%. Miners have not transformed into validators, and total block reward income (in USD) has fallen by 97\%, though transaction fees (in ETH) for Ether have increased nearly 10\%.  The Herfindahl index for the top 10 is 1,009; the network is 19\% less concentrated after the merge.  Ethereum supply growth has been deflationary since the merge. The time between consecutive blocks is now steady at 12 seconds and transactions per day are up 7.0\%. On Polygon, Matic fees rose but token fees fell.  Polygon also slows, processing 3.3\% fewer transactions per day.  Solana's fees fall by \$0.0003, and transactions per day are down 48\%.  Stablecoin transfer volumes fall on Ethereum and Polygon, but rise on Solana.

\end{abstract}
\vskip 1.25cm
\hskip 1cm \textbf{Keywords:} Ethereum; proof-of-stake; merge; cryptocurrency.
\vskip 1cm
\hskip 0.30cm \textbf{JEL Codes:} G12; G23.

\setcounter{page}{0}
\thispagestyle{empty}
\end{titlepage}


\pagebreak \newpage
\setstretch{1.05}
\section{Introduction}

The Ethereum blockchain began operation on July 30, 2015.  For more than seven years, the chain was secured by a proof-of-work (PoW) protocol. This entailed large clusters of specialized computers, known as mining networks, competing with one another to find a random number called the \textit{nonce}. The first miner to show that the hash of the nonce was below the network's difficulty level would be allowed to add the new block to the chain.  The miner would receive a block reward and collect gas fees for the transactions included in the block. As the network difficulty increased, the energy resources consumed by the miners grew ever larger.

\begin{figure}[H]
	\centering
		\caption{Electricity Consumption of the Ethereum Network}
		\label{fig:electricity_consumption}
        \begin{minipage}{0.97\linewidth}
        \begin{center}
			\includegraphics[width=0.97\textwidth]{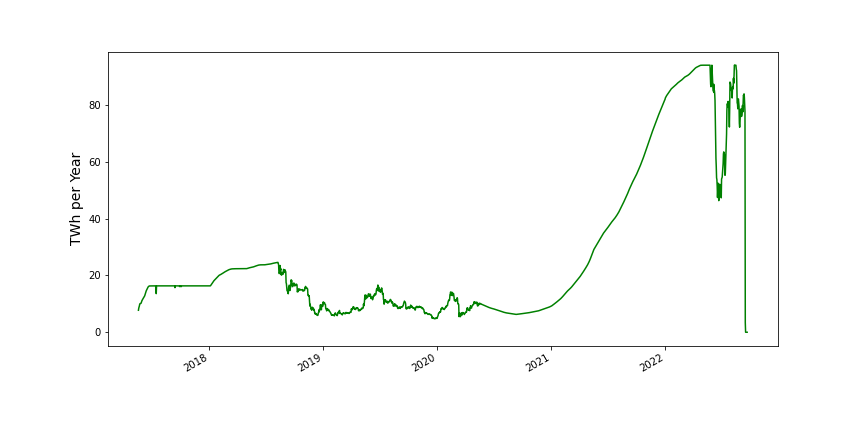} \\
		\end{center}
		\small
		\textit{Note: } The data are collected by Digiconomist, \url{https://digiconomist.net/ethereum-energy-consumption}
    \end{minipage}
\end{figure}

On August 13, 2022, the electricity consumption of the network, using the methodology of \citet{deVriesEnergy}, reached a peak, annualized at 93.975 Terra Watt Hours per year.\footnote{An alternative, more conservative methodology is used by \citet{KrauseEnergy}.} For context, this exceeds the usage by the Phillipines.\footnote{\url{https://en.wikipedia.org/wiki/List_of_countries_by_electricity_consumption}} 

Vitalik Buterin, founder of the network, had advocated for a transition to proof-of-stake (PoS) as early as 2016.\footnote{\url{https://medium.com/@VitalikButerin/a-proof-of-stake-design-philosophy-506585978d51}} This validation method requires \textit{stakers} to verify new transactions. Instead of searching randomly for the nonce, stakers place their Ethereum holdings into a smart contract as collateral. If stakers fail to fulfill their validation responsibilities--accidentally or maliciously--they can be punished by losing their staked coins. Stakers for any given block are chosen via a pseudo-random algorithm known as RANDAO.\footnote{\url{https://www.randao.org/whitepaper/Randao_v0.85_en.pdf}} \citet{BenUpgrade} provides a more comprehensive discussion about RANDAO, and \citet{ParkEth2} contains details on Ethereum's implementation of PoS.

Since PoS forgoes the energy-intensive problem solving characteristic of proof-of-work, the Ethereum transition to PoS has cut the electricity usage of the network to 0.015 Terra Watts, a 99.98\% decrease, as seen in Figure \ref{fig:electricity_consumption}.

\begin{figure}[H]
	\centering
		\caption{Ethereum Blockchain Pre and Post Merge}
		\label{fig:pre_and_post_merge}
        \begin{minipage}{0.97\linewidth}
        \begin{center}
			\includegraphics[width=0.97\textwidth]{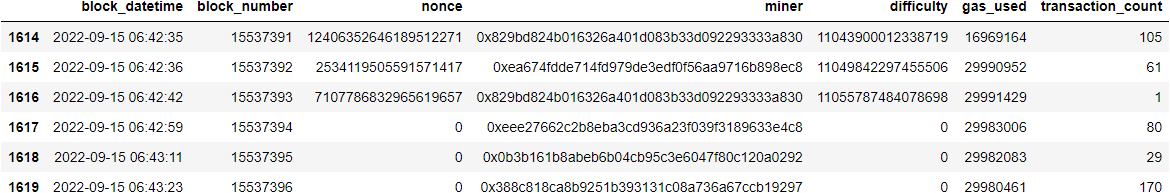} \\
		\end{center}
		\small
		\textit{Note: } Nonce and difficulty, which were critical to proof-of-work, are now empty since they are not part of proof-of-stake. The miner column is now populated by the fee recipient--the validator.  The data are from the Ethereum Mainnet which we obtain from Kaggle.
    \end{minipage}
\end{figure}

The last block mined under PoW\footnote{\url{https://etherscan.io/block/15537393}} was 15537393 by F2 Pool Old at 6:42:42 AM GMT on September 15, 2022.  It included only one ERC-721 (NFT) transaction.  The first PoS block\footnote{The block was in slot 4700013, epoch 146875.  A \textit{slot} is a time period of 12 seconds in which a validator can propose a block.  There are 32 slots in one \textit{epoch}.} 15537394 was a normal sized block\footnote{\url{https://etherscan.io/block/15537394}} with 80 transactions, but it included a 45 Ether (ETH) priority fee (\textit{tip})  We will drop the transition day, September 15, in most of our analysis, analyzing one month event windows around that date.

\section{Data and Methods}
This paper will examine the composition of the validator pool, block fees and rewards, and the network speed, on the Ethereum network.  Our data sources are the Ethereum blockchain which we obtain from Kaggle.\footnote{\url{https://www.kaggle.com/datasets/bigquery/ethereum-blockchain}}  Data from the pre-merge Beacon Chain is from Bitquery.\footnote{\url{https://explorer.bitquery.io/eth2}} We will also compare Ethereum to Polygon and Solana on fees, speed and transfer volumes.  The Polygon data are from Quicknode,\footnote{\url{https://www.quicknode.com/core-api}} and the Solana data are from its' command line interface.\footnote{\url{https://docs.solana.com/cli}}

\section{Validators}

Prior to the merge, blocks were added by the miners. For the period August 14 to September 14, 2022, we computed the number of  blocks and associated transactions that each miner completed. The top 10 miners, by blocks completed, are in Table \ref{tab:miners}. The Herfindahl index for the top ten is 1,245.  Ethermine has the largest market share at 28.6\%.

\begin{table}[H]
    \centering
    \begin{adjustbox}{width=\textwidth}
\begin{threeparttable}
  \caption{Most Active Miners Leading into the Merge}

\begin{tabular}{lrrl}
\hline
                             Miner address &  Blocks formed &  No. Trans.&     Miner names \\
\hline \hline
0xea674fdde714fd979de3edf0f56aa9716b898ec8 &          56,960 &           11,437,974 &       Ethermine \\
0x829bd824b016326a401d083b33d092293333a830 &          29,304 &            5,231,405 &          F2Pool \\
0x1ad91ee08f21be3de0ba2ba6918e714da6b45836 &          20,436 &            2,967,793 &          Hiveon \\
0x00192fb10df37c9fb26829eb2cc623cd1bf599e8 &          14,452 &            2,253,984 &         2Miners \\
0x7f101fe45e6649a6fb8f3f8b43ed03d353f2b90c &           9,859 &            1,397,143 &        Flexpool \\
0x2daa35962a6d43eb54c48367b33d0b379c930e5e &           7,004 &            1,086,420 &        Poolin 2 \\
0x52bc44d5378309ee2abf1539bf71de1b7d7be3b5 &           5,419 &            1,035,186 &        Nanopool \\
0x3ecef08d0e2dad803847e052249bb4f8bff2d5bb &           4,432 &             816,551 & Mining Pool Hub \\
0xb7e390864a90b7b923c9f9310c6f98aafe43f707 &           4,390 &             777,536 &        Unknown1 \\
0xcd458d7f11023556cc9058f729831a038cb8df9c &           3,803 &             628,777 &        Poolin 4 \\
\hline
\end{tabular}
  \label{tab:miners}
     \begin{tablenotes}
      \small
      \item \texttt{Note: } The totals are for Ethereum network blocks for the month preceding the merge, August 14 to September 14, 2022.
    \end{tablenotes}
   \end{threeparttable}%
    \end{adjustbox}
\end{table}%

\begin{figure}[H]
	\centering
		\caption{Eth 2.0 Deposit Contract}
		\label{fig:staking_contract}
        \begin{minipage}{0.97\linewidth}
        \begin{center}
			\includegraphics[width=0.97\textwidth]{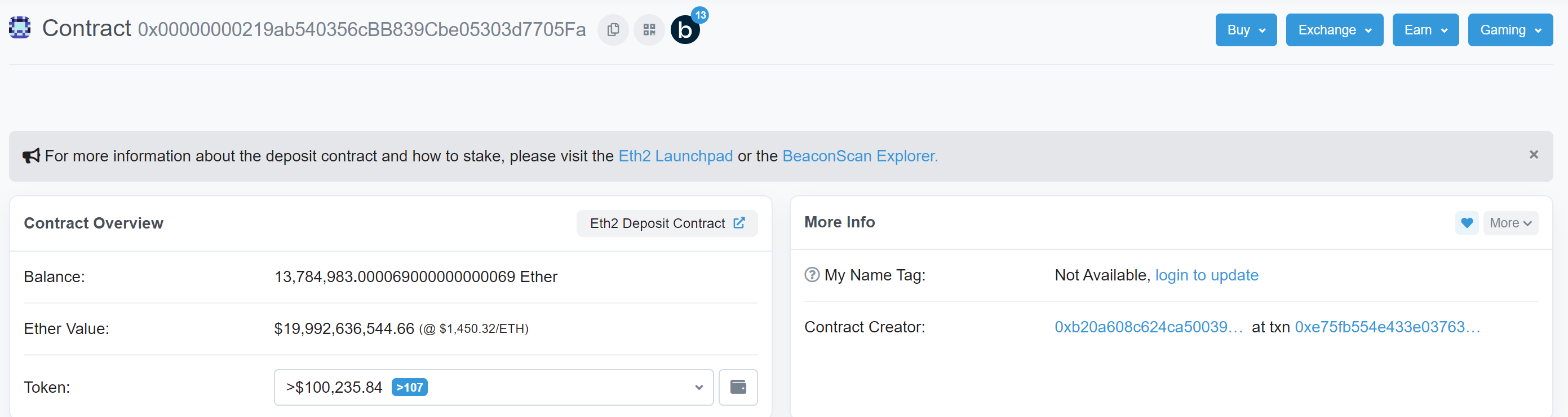} \\
		\end{center}
		\small
	    \texttt{Note: } The figure provides information about the Ethereum 2.0 Deposit Contract, created by the Beacon Contract Creator on October 14, 2020.  The image is from Etherscan.  
	\end{minipage}
\end{figure}

Blocks are now secured by \textit{validators}, participants in the Ethereum 2.0 consensus algorithm who have placed at least 32 ETH into the deposit contract\footnote{\url{https://etherscan.io/address/0x00000000219ab540356cBB839Cbe05303d7705Fa}} shown in Figure \ref{fig:staking_contract}.
A committee of at least 128 validators, selected by RANDAO, are chosen to add a block for any given slot. One participant, called the \textit{block proposer}, forms the block, a process that entails selecting and verifying a set of transactions has no failures or errors. The block then needs to be confirmed by the remaining validators, called \textit{attesters}, who check and give their vote of confidence to the block.  Finality is achieved at the \textit{checkpoint block}, the first block in the next epoch, with support from 2/3 of the staked ETH.\footnote{\url{https://ethereum.org/en/developers/docs/consensus-mechanisms/pos/finality}}

The randomness of selection means that validators must have an active system nearly 24/7; they can have some or all of their stake \textit{burned}\footnote{A term used to describe ETH that is cut loose from the network and is thus unacessible to any user.} and be removed (\textit{slashed}) from the set of validators, for failing to complete an assigned task.  As of November 1, 2022, 217 validators have been slashed,\footnote{\url{https://beaconcha.in/validators/slashings}} 24 since the merge.

\begin{figure}[H]
	\centering
		\caption{Cumulative Stake in the Ether Deposit Contract}
		\label{fig:staking_contract_supply}
        \begin{minipage}{0.97\linewidth}
        \begin{center}
			\includegraphics[width=0.97\textwidth]{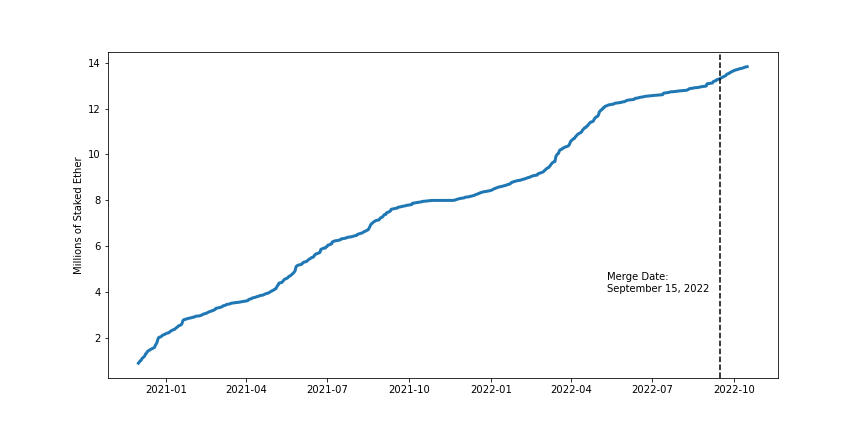} \\
		\end{center}
		\small
		\texttt{Note: } The data are from Bitquery which samples the Beacon Chain.  Deposits to the Beacon Chain staking contract began in November 2020.  Heading into the merge the total supply in the deposit contract exceeded 13.2 million Ether, and the number of validators was over 400,000.
    \end{minipage}
\end{figure}

In order to incentivize ETH holders to stake in the network, validators are compensated in two ways.  The block proposer claims the gas fees, net of burning.  Validators also receive rewards for attesting to and finalizing blocks.\footnote{\url{https://ethereum.org/en/developers/docs/consensus-mechanisms/pos/rewards-and-penalties/}} This reward is a function of the total network stake, with the return inversely proportional to the square root of the number of validators.

\citet{PintailReward} provides estimates of the returns to staking Ethereum on the Beacon Chain which draws on the Ethereum 2.0 annotated specification of \citet{BenUpgrade}.   With 401,084 validators on the merge day, this model estimates a return of 1.49 ETH over the year, an annualized yield of 4.64\%.  If the number of validators were to double, the return would fall to 1.05 ETH or just 3.28\%.

We computed the number of blocks and transactions that each validator proposed from September 16 to October 16, 2022. The top 10 validators by blocks completed can be seen in Table \ref{tab:validators}. The Herfindahl index for the top 10 is 1,009; the network is 19\% less concentrated after the merge.

\begin{table}[H]
\centering
\begin{adjustbox}{width=\textwidth}
\begin{threeparttable}
  \caption{Most Active Block Proposers Since the Merge}
\begin{tabular}{lrrl}
\hline
                      Fee receiver address &  Blocks vldtd. &  No. Trans. &       Validator name \\
\hline \hline
0xdafea492d9c6733ae3d56b7ed1adb60692c98bc5 &             45,688 &            8,809,237 &             Flashbots \\
0x388c818ca8b9251b393131c08a736a67ccb19297 &             36,753 &            5,536,598 &  Lido Execution Layer \\
0x4675c7e5baafbffbca748158becba61ef3b0a263 &             32,827 &            4,265,292 &              Coinbase \\
0xf2f5c73fa04406b1995e397b55c24ab1f3ea726c &             10,891 &            2,176,154 & bloXroute: Max Profit \\
0xebec795c9c8bbd61ffc14a6662944748f299cacf &             10,540 &            1,383,861 &             Contract1 \\
0xe688b84b23f322a994a53dbf8e15fa82cdb71127 &              9,197 &            1,221,402 &              Address1 \\
0x690b9a9e9aa1c9db991c7721a92d351db4fac990 &              7,903 &            1,694,011 &           builder0x69 \\
0x199d5ed7f45f4ee35960cf22eade2076e95b253f &              5,045 &            1,057,766 &  bloXroute: Regulated \\
0x6d2e03b7effeae98bd302a9f836d0d6ab0002766 &              4,793 &             587,538 &              Address2 \\
0xb646d87963da1fb9d192ddba775f24f33e857128 &              3,577 &             766,706 &           MEV Builder \\
\hline
\end{tabular}
  \label{tab:validators}
     \begin{tablenotes}
      \small
      \item \texttt{Note: } The totals are for one  month after the merge, September 16 to October 16, 2022.
    \end{tablenotes}
    \end{threeparttable}%
    \end{adjustbox}
\end{table}%

Moreover, there is no overlap between the pre-merge miners and the post-merge validators. Miners' comparative advantage is their computing power, but this does not give a meaningful leg-up for PoS validation. Some miners have migrated to other networks that are still using PoW. In fact, the reward for mining an Ethereum Classic or RVN blocks, which still use PoW, dropped roughly 84\% and 97\%, respectively, within a 24-hour period around the merge.\footnote{\url{https://www.coindesk.com/business/2022/09/15/ethereum-miners-are-quickly -dying-less-than-24-hours-after-the-merge/}}

The dominant validators are Flashbots, a research collective designed to mitigate the problem of maximum extractable value (MEV), a measure of the profit a network participant can make through ``their ability to arbitrarily include, exclude, or re-order transactions from the blocks they produce.''\footnote{\url{https://www.flashbots.net/}} In other words, since there is no formal regulation around the order of transactions in a block, or which transactions are included in a block, validators could choose to prioritize certain types of transactions, giving those parties systemic advantages when trading. In the proof of stake regime, Flashbots has released MEV-Boost, an open-source algorithm which sells blockspace to an open market of builders.  On September 15, 2022, 17 epochs after the merge, MEV-Boost was activated.

BioXroute\footnote{\url{https://docs.bloxroute.com/apis/mev-solution/mev-relay-for-validators}} and MEV Builder utilize MEV-Boost.\footnote{builder 0x69 utilizes two relays, Flashbots and Relayooor, \url{https://medium.com/@builder0x69}}  The four Flashbots addresses have validated 46.6\% of the blocks in the month since the merge.

While it may not be computationally intensive, validating requires a large capital commitment, a great deal of technical knowledge, and a round-the-clock connection to the network. There are, however, on- and off-chain methods for retail investors, who would otherwise be locked out due to these stringent demands, to become validators. Validation pools, such as Lido, have been a popular on-chain method of staking--in fact, as seen in Table \ref{tab:validators}, Lido is the dominant staking pool in the market. Users pool their resources into Lido by purchasing its staking token, stETH.\footnote{The hash on the Ethereum Mainnet is 0xae7ab96520de3a18e5e111b5eaab095312d7fe84.} Lido then invests the pooled assets into the staking contract, and distributes the rewards amongst the investors.

\begin{figure}[H]
	\centering
		\caption{Supply in the Lido Staking Token}
		\label{fig:lido_supply}
        \begin{minipage}{0.97\linewidth}
        \begin{center}
			\includegraphics[width=0.97\textwidth]{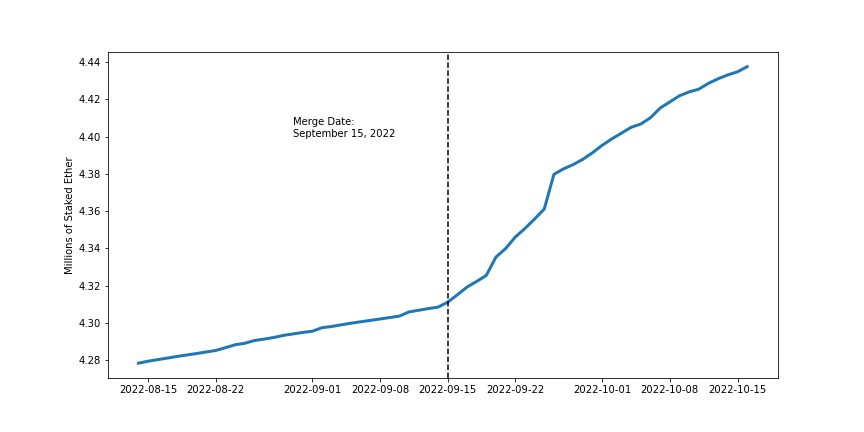} \\
		\end{center}
		\small
		\texttt{Note: } The plot is the cumulative supply change in the Lido staking token from August 14 to October 16, 2022.
    \end{minipage}
\end{figure}

There are also off-chain solutions for retail staking, namely through centralized exchanges. As seen on Table \ref{tab:validators}, Coinbase is the second largest staker in part due to a service that mirrors Lido's on-chain pooling; Coinbase users can offer any amount of ETH to be pooled and staked in the deposit contract.  Other centralized exchanges, like Binance, offer similar services; Kraken offered such a service for the duration of this sample, but have since shut down user stalking pools.  Since daily issuance of new ETH to the validators is a function of the total amount of ETH staked, tracking the dominant validators may be important insofar as understanding the growth in ETH supply.

Validators can't withdraw ETH from the staking contract until the Shanghai upgrade. At the time of the Merge, it was anticipated that the upgrade would take six to eighteen months, and it is now on pace for the lower end of that range, March 2023.\footnote{https://www.bloomberg.com/news/articles/2023-01-27/crypto-s-hottest-trade-risks-spurring-another-leverage-bubble\#xj4y7vzkg} 

\section{Block Income and Creation Speed}
Historically, income for miners has come from two sources: block rewards for completing a new block on the chain and transaction fees from on-chain transfers.  Since August 2021, after the EIP-1559 upgrade,\footnote{\url{https://notes.ethereum.org/@vbuterin/eip-1559-faq}} base fees from transfers have been burned. A miner's block income, prior to the merge, consisted of these three parts shown in (\ref{eq:block_income}):

\begin{equation} \label{eq:block_income}
    \text{Block Income = Block Reward + Transaction Fees - Burnt Gas}
\end{equation}

Block reward is the number of new ETH minted and given to the party responsible for successfully adding a block to the chain; transaction fees, include the gas, both base and priority fees, from the transactions in the block; burnt gas for each transaction is the base fee in ETH set by the network protocol.

\begin{figure}[H]
	\centering
		\caption{Total Block Income in Ether}
		\label{fig:block_rewards}
        \begin{minipage}{0.97\linewidth}
        \begin{center}
			\includegraphics[width=0.97\textwidth]{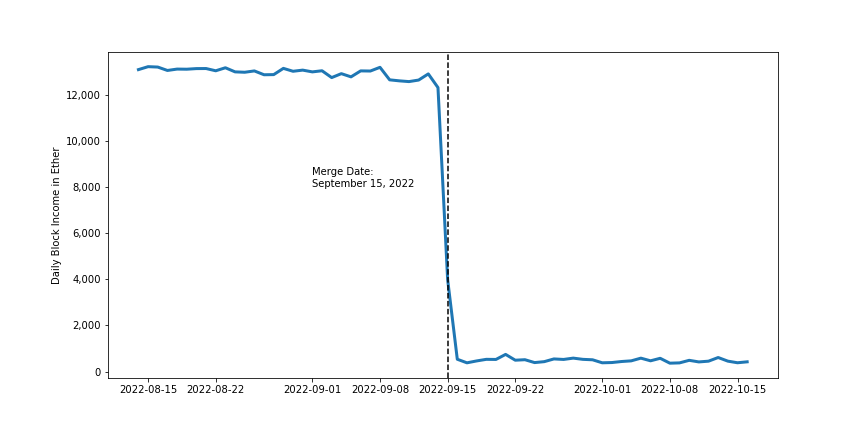} \\
		\end{center}
		\small
		\texttt{Note: } The data are from the Ethereum Mainnet which we obtain from Kaggle.  The sample period is August 14 to October 16, 2022.
    \end{minipage}
\end{figure}

Since the St. Petersburg upgrade in 2019, miners received two ETH for each block they completed. However, the transition to PoS removed this reward altogether. Miner and validators' total block income before and after the event is plotted in Figure \ref{fig:block_rewards}.

Table \ref{tab:block_income} decomposes the decline in block income.  There is a slight, statistically insignificant rise in transaction fees for transfers, but it is largely offset by a higher burn rate. The loss of a block completion bonus is the main drag on block income. Validators' total block income is now only 3\% of what miners used to earn.
\begin{table}[H]
\centering
\begin{threeparttable}
  \caption{Daily Block Income Before and After the Merge}
\begin{tabular}{lrrrr}
    \hline
        Measure & Pre- & Post- & t-stat & p-val  \\ 
    \hline \hline
  Block Tx Fees   (Ether)  & 1,675 &   1,889 &    1.41 & 0.1643 \\
   Block Burn Fees (Ether) & 1,207 &   1,413 &    1.41 & 0.1648 \\
Total Block Income (Ether) & 12,462 &    479 & -300.95 & 0.0000 \\
 Total Block Income (USD)  & \$20,732,030 & \$638,820 &  -61.52 & 0.0000 \\
\hline
\end{tabular}
  \label{tab:block_income}
     \begin{tablenotes}
      \small
      \item \texttt{Note: } The totals are for the month prior to the merge, August 14 to September 14, 2022 and for the month after, September 16, to October 16, 2022.
    \end{tablenotes}
    \end{threeparttable}%
\end{table}%
   
The removal of the block completion reward has also greatly slowed the minting of new ETH. Since the St. Petersburg Upgrade,\footnote{The updates went live at block 7,280,000 on February 28, 2019: \url{https://cointelegraph.com/news/ethereums-constantinople-st-petersburg-upgrades-have-been-activated}} two new ETH were minted as a reward for each completed block; this served as the main driver of ETH supply expansion. Combined with a smaller issuance for ommer blocks\footnote{\url{https://ethereum.org/en/glossary/ommer}} and validators on the Beacon Chain,\footnote{\url{https://ethereum.org/en/upgrades/beacon-chain/}} there was about 5.5 million new ETH minted annually.\footnote{\url{https://ethereum.org/en/upgrades/merge/issuance/}} Since the merge, the two ETH per transaction reward on the execution layer (PoW) are no longer being produced. There is now roughly 1,600 new ETH minted each day, which is distributed to the validators in the network.  
The burn rate can, in principal, exceed the issuance rate, resulting in blocks that deflate  total ETH supply.  Finally, ETH continues to be staked, removing it from the circulating supply.  

\begin{figure}[H]
	\centering
		\caption{Ethereum Supply}
		\label{fig:ether_supply}
        \begin{minipage}{0.97\linewidth}
        \begin{center}
			\includegraphics[width=0.97\textwidth]{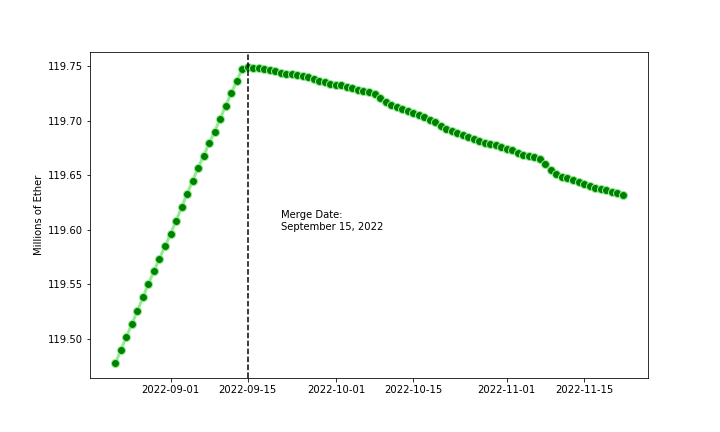} \\
		\end{center}
		\small
		\texttt{Note: } The sample is from August 14 to October 16, 2022.  The data are from Coinmetrics.
    \end{minipage}
\end{figure}

These factors explain why, as we show in Figure \ref{fig:ether_supply}, that ethereum supply has been deflationary since the merge.

\subsection{Transaction fees: Ether transactions}

As \citet{DonezFees} note, miners generally received only the base fee of 21,000 units of gas for ETH transfers between two wallets.  The dollar price of the transaction can vary quite widely because: (1) the ETH price of a unit of gas deviates with network congestion and (2) the dollar price of ETH, until recently, was increasing steadily.

\begin{figure}[H]
	\centering
		\caption{Median Fees for Ether Transactions}
		\label{fig:eth_median_fees}
        \begin{minipage}{0.97\linewidth}
        \begin{center}
			\includegraphics[width=0.97\textwidth]{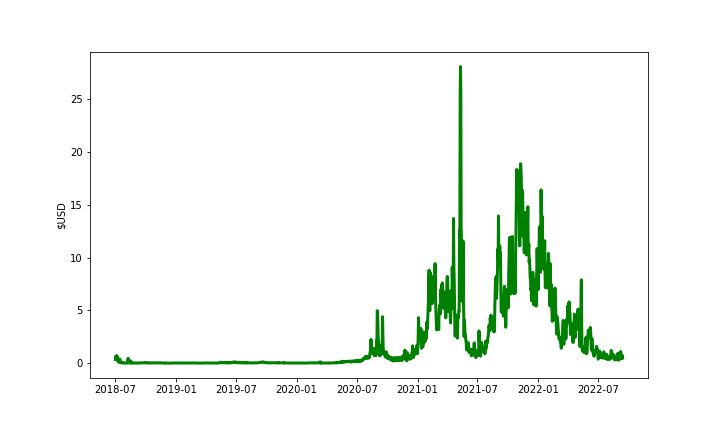} \\
		\end{center}
		\small
		\texttt{Note: } The sample is July 1, 2018 to September 14, 2022
    \end{minipage}
\end{figure}

The median fee in Figure \ref{fig:eth_median_fees} peaked at over \$28 on May 11, 2021. Following a temporary dip, prices remained over \$10 through November 2021, but have fallen since.  Going into the merge, the median price was \$0.502.

As seen in Table \ref{tab:eth_fees}, gas price and gas used for ETH transactions have seen statistically significant increases since the merge. That said, given the strong decline in ETH/USD during the event window, there has actually been a slight decrease in the dollar cost of fees, but this change is not statistically significant.

\begin{table}[H]
\centering
\begin{threeparttable}
  \caption{Ether Transaction Fees Before and After the Merge}
\begin{tabular}{lrrrr}
    \hline
            Measure & Pre- & Post- & t-stat & p-val  \\ 
    \hline \hline
Gas Price (Gwei)  &  17.3279 &  22.6959 &    2.76 & 0.0070 \\
        Gas Used $\cdot 10^{4}$  & 6.8570 & 7.3321 &    3.85 & 0.0002 \\
    Fees in Ether $\cdot 10^{-3}$  & 1.1719 & 1.2872 &    3.60 & 0.0005 \\
     Fees in USD  &     \$1.9453 &     \$1.7031 &   -1.62 & 0.0906\\
        \hline
    \end{tabular}
  \label{tab:eth_fees}
     \begin{tablenotes}
      \small
      \item \texttt{Note: } The averages are for the month prior to the merge, August 14 to September 14, 2022 and for the month after, September 16 to October 16, 2022. 
    \end{tablenotes}
    \end{threeparttable}%
\end{table}%

\subsection{Transaction fees: Ethereum tokens}

\begin{figure}[H]
	\centering
		\caption{Median Fees for Ether Token Transactions}
		\label{fig:eth_median_token_fees}
        \begin{minipage}{0.97\linewidth}
        \begin{center}
			\includegraphics[width=0.97\textwidth]{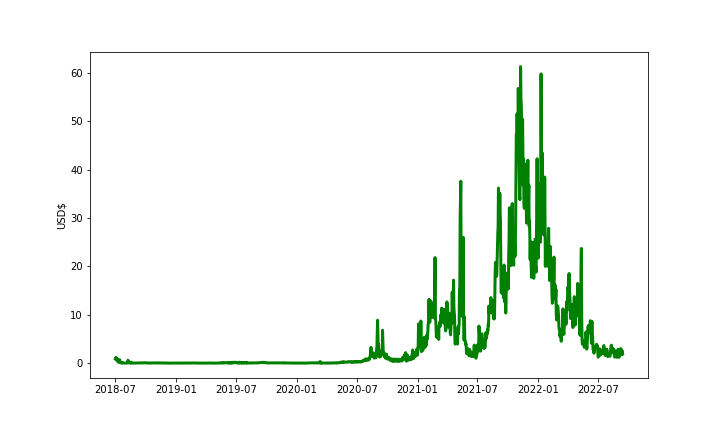} \\
		\end{center}
		\small
		\texttt{Note: } The sample is July 1, 2018 to September 14, 2022
    \end{minipage}
\end{figure}

By construction, ERC-20 tokens live in smart contracts, which introduce more complexity to parse. Thus, fees have typically been higher for token transfers. As seen in Figure \ref{fig:eth_median_token_fees}, token fees peaked at over \$61 on November 9, 2021 and medians were still above \$20 through May 2021.

We analyze the impact of the merge on ERC-20 transactions in Table \ref{tab:token_fees}.

\begin{table}[H]
\centering
\begin{threeparttable}
  \caption{ERC-20 Token Transaction Fees Pre- and Post-Merge}
\begin{tabular}{lrrrr}
    \hline
        Measure & Pre- & Post- & t-stat & p-val  \\ 
\hline \hline
Gas Price (Gwei)  &  18.8273 &   18.3042 &   -0.35 & 0.7246 \\
        Gas Used \(\cdot10^{5}\)  & 1.1104 & 1.1443 &    1.86 & 0.0670 \\
    Fees in Ether \(\cdot10^{-3}\) & 1.9458 & 1.9706 &    0.14 & 0.8868 \\
     Fees in USD  &     \$3.2387 &     \$2.6106 &   -2.56 & 0.0131 \\
    \hline
    \end{tabular}
  \label{tab:token_fees}
     \begin{tablenotes}
      \small
      \item \texttt{Note: } The averages are for the month prior to the merge, August 14 to September 14, 2022 and for the month after, September 16 to October 16, 2022.
    \end{tablenotes}
    \end{threeparttable}%
\end{table}%
Gas prices are down, gas used on token transactions is up, resulting in a small increase in fees in Ether terms.  The decline in ETH/USD leads to the only statistically significant change, a decline in the dollar cost of transfer fees of \$0.62.

\subsection{Block creation speed}
Prior to the merge, blocks were added when a miner group ``won the race'' to find the nonce. As a result, block creation speed would vary.  Except on the rare occasions that a block is skipped, validators produce a new block every twelve seconds.  Given this faster speed, transactions per day are up by 75,000 on average.

\begin{table}[H]
\centering
\begin{threeparttable}
  \caption{Daily Average Transactions Before and After the Merge}
\begin{tabular}{lrrrr}
\toprule
                Measure &   Pre- &  Post- &  t-stat &  p-val \\
\hline \hline
Transactions per day  & 1,073,508 & 1,148,750 &    5.29 &    0.0 \\
\hline
\end{tabular}
  \label{tab:chain_speed}
     \begin{tablenotes}
      \small
      \item \texttt{Note: } The averages are for the month prior to the merge, August 14 to September 14, 2022 and for the month after, September 16 to October 16, 2022.
    \end{tablenotes}
    \end{threeparttable}%
\end{table}

\begin{figure}[H]
	\centering
		\caption{Transactions per Day}
		\label{fig:block_speed}
        \begin{minipage}{0.97\linewidth}
        \begin{center}
			\includegraphics[width=0.97\textwidth]{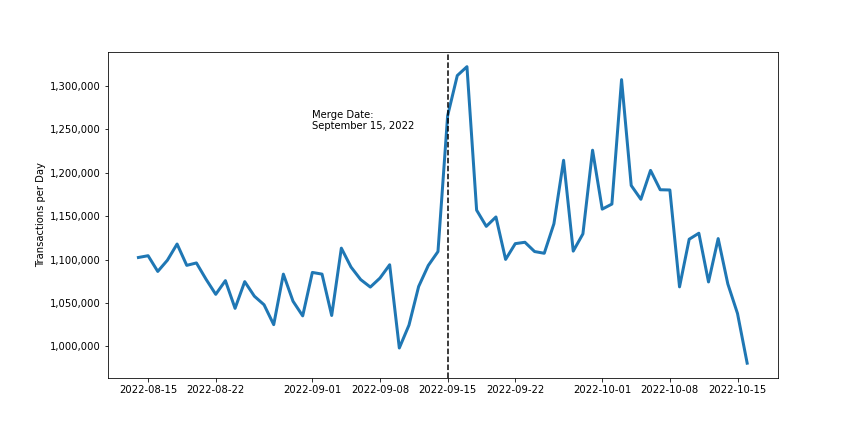} \\
		\end{center}
		\small
		\texttt{Note: } The sample is August 14 to October 16, 2022.
    \end{minipage}
\end{figure}

\begin{figure}[H]
	\centering
		\caption{Skipped Blocks per Day}
		\label{fig:skips_per_day}
        \begin{minipage}{0.97\linewidth}
        \begin{center}
			\includegraphics[width=0.97\textwidth]{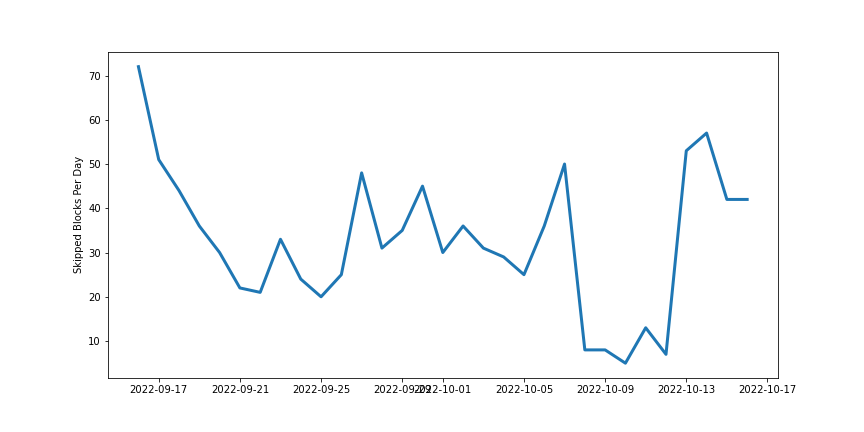} \\
		\end{center}
		\small
		\texttt{Note: } The sample is September 16 to October 16, 2022.
    \end{minipage}
\end{figure}

In fact, as seen in Figure \ref{fig:skips_per_day}, there were only 32.5 blocks per day, on average, that were not created exactly 12 seconds after the preceding block. All these blocks were created 24 seconds afterwards. This consistency occurs by design: blocks are now added by chosen validators into specific slots, each 12 seconds apart. The only reason why slots are ``skipped,'' is due to a failed block creation.

Increased and more stable block formation speed could allow for more on-chain activity. In theory, there seems to be nothing preventing an even faster rate. That said, there may be an optimal balance between block creation speed and the number of transactions per block, although lower latency networks, such as Solana, provide much higher transaction volumes. 

\section{Polygon}

Polygon is a layer 2 scaling solution on the Ethereum network. While it exists alongside the original chain, it creates a separate chain that is faster, maintaining higher transactions per second, and lowers fees per transaction.

We analyze both Matic transactions, the utility token on Polygon, and all other tokens in Table \ref{tab:polygon_fees}.

\begin{table}[H]
\centering
\begin{threeparttable}
  \caption{Transaction Fees on Polygon Before and After the Merge}
\begin{tabular}{lrrrr}
\hline
   Measure              & Pre-    & Post-    & t-stat & p-val  \\
\hline \hline
Matic Fees in USD & \$0.0232 & \$0.0242 &    0.28 &   0.78 \\
Token Fees in USD & \$0.1333 & \$0.1244 &   -0.39 &   0.70 \\
\hline
\end{tabular}
  \label{tab:polygon_fees}
     \begin{tablenotes}
      \small
      \item \texttt{Note: } The averages are for Polygon transactions in the month before the merge, August 14 to September 14, 2022, and the month after, September 16 to October 16, 2022. The Polygon data are from Quicknode, \url{https://www.quicknode.com/core-api}.
    \end{tablenotes}
    \end{threeparttable}%
\end{table}%
The merge's cross network impact lowers Polygon's Matic fees and raises token fees on the Mainnet.  Neither change is statistically significant.

We will next analyze the network speed around the merge, as measured in Table \ref{tab:polygon_speed}.

\begin{table}[H]
\centering
\begin{threeparttable}
  \caption{Network Speed Measures for Polygon}
\begin{tabular}{lrrrr}
\toprule
          Measure &     Pre- &    Post- &  t-stat &  p-val \\
\hline \hline
Transactions per day  & 2,821,632 & 2,729,498 &   -1.49 &   0.14 \\
\bottomrule
\end{tabular}
  \label{tab:polygon_speed}
     \begin{tablenotes}
      \small
      \item \texttt{Note: } The averages are for Polygon transactions in the month before the merge, August 14 to September 14, 2022, and the month after, September 16 to October 16, 2022.
    \end{tablenotes}
    \end{threeparttable}%
\end{table}%

The number of transactions per day fell by about 98,000 (3\%), but this change is not statistically significant.

\section{Solana}
Solana is a open source, permissionless blockchain that relies on proof-of-history.\footnote{\url{https://docs.solana.com/introduction}} It has substantially lower fees than either the Mainnet or Polygon.  In Table \ref{tab:solana_token_fees}, we report a small reduction in fees, significant at the 1\% level.

\begin{table}[H]
\centering
\begin{threeparttable}
  \caption{Token Transaction Fees on Solana Before and After the Merge}
\begin{tabular}{lrrrr}
\hline
       Measure &         Pre- &        Post- &   t-stat &    p-val \\
\hline \hline
Token Fees in USD & \$0.0011 & \$0.0008 &   -3.32 &    0.001 \\
\hline
\end{tabular}
 \label{tab:solana_token_fees}
     \begin{tablenotes}
      \small
      \item \texttt{Note: } The averages are for token transactions on  Solana in the month before the merge, August 14 to September 14, 2022, and the month after, September 16 to October 16, 2022.
    \end{tablenotes}
    \end{threeparttable}%
\end{table}%

As for speed, Table \ref{tab:solana_network_speed} shows that the network, prior to the merge, processes almost 40 times as many transactions as Ethereum and 15 times as many as Polygon.

\begin{table}[H]
\centering
\begin{threeparttable}
  \caption{Solana Network Speed}
\begin{tabular}{lrrrr}
\hline
         Measure &      Pre- &     Post- &  t-stat &  p-val \\
\hline \hline
Transactions per day & 42,400,730	&	22,127,360
 &    -6.89 &  0.000 \\
\hline
\end{tabular}
 \label{tab:solana_network_speed}
     \begin{tablenotes}
      \small
      \item \texttt{Note: } The averages are for 25,000 slot slices from each epoch in the month before and after the merge. We gather the data from the Solana command line interface, \url{https://docs.solana.com/cli}
    \end{tablenotes}
    \end{threeparttable}%
\end{table}%

Solana slows by 48\% after the merge.  This appears to  be related to issues to network outages on Solana.\footnote{Coin Telegraph, August 13, 2022, \url{https://cointelegraph.com/news/ominous-solana-technicals-hint-at-sol-price-crashing-35-by-september}}  Despite this, Solana remains 19 times faster than Ethereum and eight times faster than Polygon.

\section{Transfer Volumes}
We analyze transfer volume on Ethereum, Polygon, and Solana in two major stablecoins, USD Coin and Tether.  We choose these tokens because they are issued natively on the three platforms,\footnote{USD Coin: \url{https://www.circle.com/en/usdc-multichain/}; Tether: \url{https://tether.to/en/transparency/}} are traded actively, and have a stable value.  Ethereum dominates the other two platforms prior to the merge with 89\% of the transfer volume in USDC and 76\% of the volume in USDT.

\begin{table}[H]
\centering
\begin{threeparttable}
  \caption{USDC Transfer Volume}
\begin{tabular}{lrrrr}
\hline
            USD Coin ADV bn. &   Pre &  Post &  t-stat &  p-val \\
\hline \hline
Ethereum & 13.81 & 11.02 &   -1.75 &  0.09 \\
Polygon & 0.63 & 0.47 &   -1.88 &   0.07 \\
Solana & 1.00 & 1.53 &    1.65 &   0.12 \\
\hline
\end{tabular}
    \label{tab:usdc_volume}
     \begin{tablenotes}
      \small
      \item \texttt{Note: } The averages are for one month samples before and after the merge.
    \end{tablenotes}
    \end{threeparttable}%
\end{table}%

Solana's USDC volume rises more than 500 million per day, and it regains market share after the merge, rising from 6\% to 11\%.

\begin{table}[H]
\centering
\begin{threeparttable}
  \caption{USDT Transfer Volume}
\begin{tabular}{lrrrr}
\hline
            Tether ADV bn. &   Pre &  Post &  t-stat &  p-val \\
\hline \hline
Ethereum &  2.97 &  2.75 &   -0.97 &  0.34 \\
Polygon & 0.16 & 0.14 &   -1.21 &   0.23 \\
Solana & 0.80 & 1.08 &    1.52 &   0.14 \\
\hline
\end{tabular}
 \label{tab:usdt_volume}
     \begin{tablenotes}
      \small
      \item \texttt{Note: } The averages are for one month samples before and after the merge.
    \end{tablenotes}
    \end{threeparttable}%
\end{table}%

Solana's USDT volume average is again the only to rise after the merge. The transfer volume increase of 280 million per day raises its' market share to 27\% after the merge.
  
\section{Conclusion}

The main social benefit from the merge has been the massive decrease in the amount of energy used by the network; continuation of proof-of-work was likely unsustainable.

Surprisingly though, transaction fees in Ether for both Ether-only and ERC-20 transactions have risen since the merge.  We summarize our findings in Table \ref{tab:fee_summary}. Declining prices for the utility tokens lower network fees in USD, except for Matic transactions on Polygon.

\begin{table}[H]
\centering
\begin{threeparttable}
  \caption{Fees in USD on Ethereum, Polygon and Solana}
    \centering
    \begin{tabular}{llll}
    \hline
        & Pre- & Post-   \\ \hline \hline
     Ether Transactions  &     \$1.9453 &     \$1.7031 \\
        Ethereum Tokens  &     \$3.2387 &     \$2.6106\\ 
        Polygon Matic Transactions & \$0.0232 & \$0.0242 \\ 
        Polygon Tokens & \$0.1333 & \$0.1244 \\ 
        Solana Tokens & \$0.0011 & \$0.0008\\ 
\hline
    \end{tabular}
 \label{tab:fee_summary}
     \begin{tablenotes}
      \small
      \item \texttt{Note: } The averages are for one month samples before and after the merge.
    \end{tablenotes}
    \end{threeparttable}%
\end{table}

As summarized in Table \ref{tab:speed_summary}, the speed of activity rose slightly on Ethereum and fell slightly on Polygon.  Solana has slowed by almost 50\% which may be attributable to major network outages on September 30 and October 1, 2022.\footnote{https://status.solana.com/uptime?page=2}

\begin{table}[H]
\centering
\begin{threeparttable}
  \caption{Speed on Ethereum, Polygon and Solana}
    \centering
    \begin{tabular}{lrrrr}
    \hline
         Transactions per day &      Pre- &     Post-  \\
\hline \hline
Ethereum & 1,073,508 & 1,148,750 \\
Polygon & 2,821,632 & 2,729,498 \\
Solana & 42,400,730	&	22,127,360 \\
\hline
    \end{tabular}
 \label{tab:speed_summary}
     \begin{tablenotes}
      \small
      \item \texttt{Note: } The event window is the same as in Table \ref{tab:fee_summary}.
    \end{tablenotes}
    \end{threeparttable}%
\end{table}

The transition from proof-of-work to proof-of-stake has also reshaped the entire Ethereum landscape. The composition of block creators has entirely changed; there is not one miner from the PoW regime that is now proposing or attesting to blocks as a validator under the PoS regime. While gas transaction fees for network participants have grown since the merge, so has the amount of ETH burned. Block rewards were many times larger than the daily variable sum, distributed to the validators.  Because of the loss of block rewards, increased burn rates, and more staked Ether, the total circulating supply is now deflationary.  

There are many open questions remaining.  What will be the long run return to staking? Two ongoing factors are: when will validators finally be able to withdraw stalked ETH;\footnote{\url{https://www.bloomberg.com/news/articles/2022-09-14/-illiquidity-risk-is-a-side-effect-of-ethereum-crypto-upgrade}} 216 validators have already been slashed for improper actions as validators;\footnote{\url{https://www.blocknative.com/blog/an-ethereum-stakers-guide-to-slashing-other-penalties}} how attractive will this investment be in the long run?

Will decentralized staking systems like Lido dominate centralized exchanges like Coinbase? Will Ethereum gas fees eventually fall given the reduction in energy needed to secure the network? Will other proof-of-work networks look to transition to proof-of-stake or other less energy intensive models of block formation? Finally, will sharding, which effectively makes individual nodes into their separate blockchains,\footnote{https://ethereum.org/en/upgrades/sharding/.  It is projected to arrive sometime in 2023.} enable Ethereum to achieve latencies and fees similar to Solana?

Lastly, major policy questions are still unresolved in the digital asset space. Specifically, Chairman Gensler of the SEC has suggested that staking may qualify certain blockchains as securities, as per the Howey Test,\footnote{ Paul Kiernan and Vicky Ge Huang,
Ether’s New ‘Staking’ Model Could Draw SEC Attention. \textit{Wall Street Journal}, September 16, 2022.} though Ethereum is currently classified as a commodity by the Commodity Futures Trading Commission (CFTC).\footnote{The CFTC asserts this in their filing again Sam Bankman-Fried: ''A digital asset is anything that can be stored and transmitted electronically and has
associated ownership or use rights. Digital assets ... function as
mediums of exchange, units of account, and/or stores of value. Certain digital assets are
“commodities,” including...ether (ETH)... as defined under
Section 1a(9) of the Act, 7 U.S.C. § 1a(9).} Kraken has recently discontinued its Staking-As-A-Service program as part of a \$30 million settlement with the SEC.\footnote{\url{https://www.sec.gov/news/press-release/2023-25}} Gensler has made it clear\footnote{\url{https://www.sec.gov/news/statement/gensler-statement-custody-021523}} that centralized exchanges are not ``qualified custodians," so this action against Kraken could potentially be applied more broadly. If the custodial rules were also applied to on-chain assets like Lido, nearly 50\% of staked Ether could be impacted.

\pagebreak
\begingroup
\setstretch{1.25}
\printbibliography
\endgroup

\end{document}